\documentclass[
 aip,
 prl,
 amsmath,amssymb,
]{revtex4-1}

\usepackage{gensymb}
\usepackage{graphicx}

\begin{document}

\title{Thermal switching of indirect interlayer exchange in magnetic multilayers}
 
\author{D.~M.~Polishchuk}
 \email{dpol@kth.se.}
 \affiliation{Nanostructure Physics, Royal Institute of Technology, Stockholm, Sweden}%

\author{Yu.~O.~Tykhonenko-Polishchuk}
\author{A.~F.~Kravets}
 \affiliation{Nanostructure Physics, Royal Institute of Technology, Stockholm, Sweden}
 \affiliation{Institute of Magnetism, NAS of Ukraine, Kyiv, Ukraine}
 
\author{V.~Korenivski}%
 \affiliation{Nanostructure Physics, Royal Institute of Technology, Stockholm, Sweden}%
 
\date{\today}

\begin{abstract}

We propose a magnetic multilayer layout, in which the indirect exchange coupling (IEC also known as RKKY) can be switched on and off by a slight change in temperature. We demonstrate such on/off IEC switching in a Fe/Cr/FeCr-based system and obtain thermal switching widths as small as 10--20~K, essentially in any desired temperature range, including at or just above room temperature. These results add a new dimension of tunable thermal control to IEC in magnetic nanostructures, highly technological in terms of available materials and operating physical regimes.

\end{abstract}

\maketitle

\section{Introduction}

Since the discovery of the indirect exchange coupling \cite{r1} in magnetic multilayers, followed by the discovery of the related effect of giant magnetoresistance \cite{r2,r3}, numerous material systems both metallic and semiconducting were reported to poses it, among which the pioneering Fe/Cr/Fe and Co/Cu/Co structures have been the most studied (see refs.~\cite{r4,r5} and references therein). The distinctive feature of the IEC is that the strength as well as sign of the coupling are functions of the thickness of the nonmagnetic spacer \cite{r6}. Such oscillatory character results in either parallel or antiparallel mutual orientation of the magnetic moments of the neighboring ferromagnetic layers as the spacer thickness is varied. The mechanism of IEC is closely related to the Ruderman-Kittel–Kasuya-Yosida (RKKY) interaction acting between magnetic impurities in a non-magnetic host \cite{r7,r8,r9}. Even though the RKKY approach may not be fully applicable to the IEC in transition metal multilayers \cite{r4,r10}, starting with the early models \cite{r11,r12,r13} it generally captures well the physics involved. Most of the experimental data on IEC to date have been successfully described in terms of a more general quantum-well approach \cite{r14,r15}, where the spin-dependent density of states of the nonmagnetic spacer depends on the relative orientation of the ferromagnetic layers, which favors either parallel (ferromagnetic, FM) or antiparallel (antiferromagnetic, AFM) alignment of the respective magnetic moments. The period of the IEC oscillation depends on the properties of the spacer’s Fermi surface, while the amplitude as well as the phase may be affected by the interface roughness, interdiffusion, and related effects.
 
Variation in temperature has a relatively minor effect on the IEC in systems with metallic \cite{r16}  or semiconducting spacer layers \cite{r17}. In the first case, a slight weakening of the coupling strength with increasing temperature is well explained by thermal broadening of the Fermi edge \cite{r13}. In the second case, increasing temperature has the inverse effect \cite{r17} due to increased thermal population of the conduction band of the semiconducting spacer \cite{r10,r18}. A number of reports have shown a stronger temperature dependence than that predicted by theory \cite{r19,r20,r21,r22,r23}. The measured change in the IEC strength was up to 75~\% over a 300~K interval, which was stronger than the predicted behavior \cite{r10}, though still far from being technologically interesting for thermo-magnetic control of nanodevices.   

\begin{figure*}[t!]
\includegraphics[width=140mm]{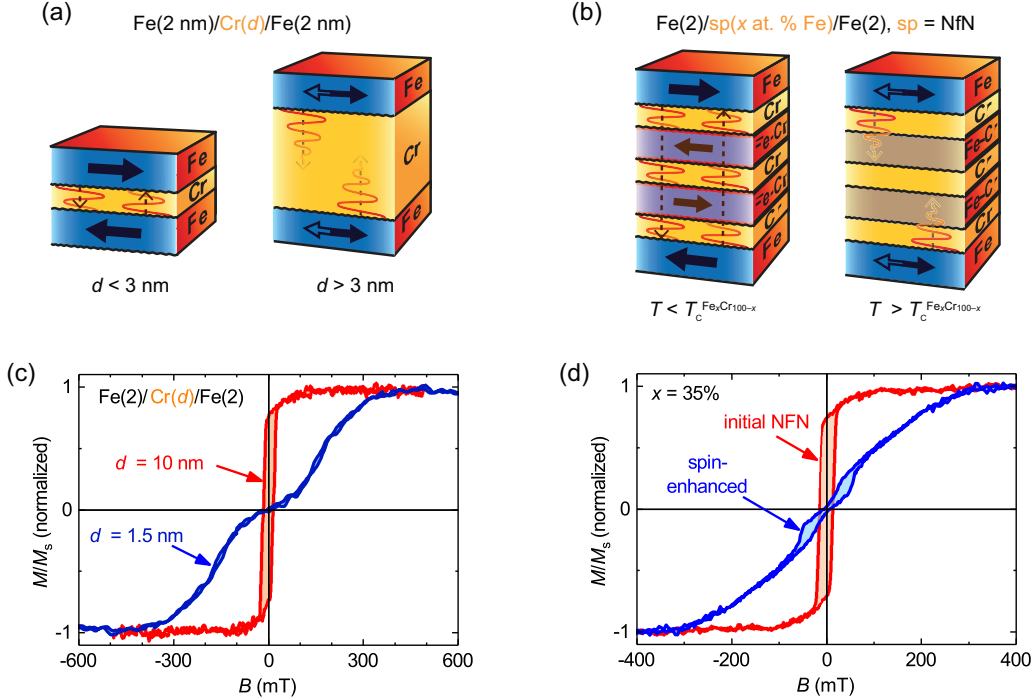}
\caption{(a) Illustration of reference samples Fe(2)/Cr($d$)/Fe(2) without IEC (right panel) and with IEC (left panel). Solid arrows denote the magnetic moments of the Fe layers subject to IEC; two-directional arrows denote mutually independent orientation of the Fe magnetic moments in the absence of IEC. (b) Modified multilayer layout with the outer Fe layers RKKY-coupled antiferromagnetically via two weakly ferromagnetic Fe$_x$Cr$_{100-x}$ alloy layers, suitably placed within the composite spacer of NfN-type. Right (left) panel shows the corresponding magnetic layout above (below) the Curie temperature of the Fe$_x$Cr$_{100-x}$ layers. The same picture holds for the structures with ultrathin Fe at the Cr/Fe-Cr interfaces (spin-enhanced interfaces). (c) Room temperature M-B loops for reference Fe(2)/Cr($d$)/Fe(2) samples with $d$ = 1.5 and 10~nm and (d) for NfN-samples with and without spin-enhancement, respectively.}
\label{fig_1}
\end{figure*}

Another approach towards thermal control of IEC was to change the sign of the RKKY interaction via Y spacers in Ga/Y/Tb tri-layers \cite{r24} and via Pt spacers in perpendicularly magnetized Co/Pt/[Co/Py]$_\text{n}$ multilayers \cite{r25}. In both systems, the observed effect was explained as due to thermal changes in the magnetization of the respective soft outer magnetic layers (Co and Ga), affecting the phase of the IEC. The reported changes in the IEC were rather weak, however, of the order of 1~mT, well within the coercivity of the ferromagnetic layers used and, as a result, no clear parallel-to-antiparallel thermal switching was obtained. 

Here we report on a new multilayer design for efficient thermal switching of IEC, illustrated in fig.~\ref{fig_1}, in which a diluted ferromagnetic 3d-metal-alloy layer within a composite spacer is tailored to have its Curie point ($T_\text{C}$) at a desired transition temperature, and is spaced by nonmagnetic layers such as to either transmit or not transmit AFM-RKKY interaction, in its ferromagnetic (below $T_\text{C}$) or paramagnetic (above $T_\text{C}$) state, respectively. We demonstrate sharp on/off thermal switching of a rather strong IEC, with 10--20~K transition widths and 0.1--1~T switching field strengths, respectively. This, combined with a broad choice of suitable materials and wide tuneability as regards the physical and operating parameters, makes the demonstrated system highly technological.

\section{Samples and Experiment}

Three series of multilayers were investigated: (i) reference series Fe(2)/Cr($d$)/Fe(2), with $d~=$ 1--10~nm; (ii) series with composite spacers Fe(2)/[Cr(1.5)/Fe$_x$Cr$_{100-x}$(3)]$_{\times2}$/ Cr(1.5)/Fe(2); and (iii) series with spin-enhanced composite spacers Fe(2)/[Cr(1.5)/Fe(0.25)/ Fe$_x$Cr$_{100-x}$(3)/ Fe(0.25)]$_{\times2}$/Cr(1.5)/Fe(2). The thickness in parentheses are in 'nm'.  The series with composite and spin-enhanced composite spacers had three samples in each, with $x$ = 30, 35 and 40 at.~\% Fe in the diluted ferromagnetic Fe$_x$Cr$_{100-x}$ layers.

The multilayers were deposited at room temperature onto Ar pre-etched un-doped Si (100) substrates by dc-magnetron sputtering. Layers of dilute Fe$_x$Cr$_{100-x}$ binary alloys of varied composition were deposited using co-sputtering from separate Fe and Cr targets. The composition of the Fe$_x$Cr$_{100-x}$ layers was controlled by setting the corresponding deposition rates of the individual Fe and Cr components, with relevant calibrations obtained by subsequent thickness profilometry. The in-plane magnetic measurements were performed in the temperature range of 20–120 $^{\circ}$C using a vibrating-sample magnetometer equipped with a high-temperature furnace (Lakeshore Inc.) as well as a magneto-optical Kerr effect magnetometer equipped with an optical cryostat (Oxford Instr.).

\section{Results and Discussion}

\begin{figure*}
\centering
\includegraphics[width=140mm]{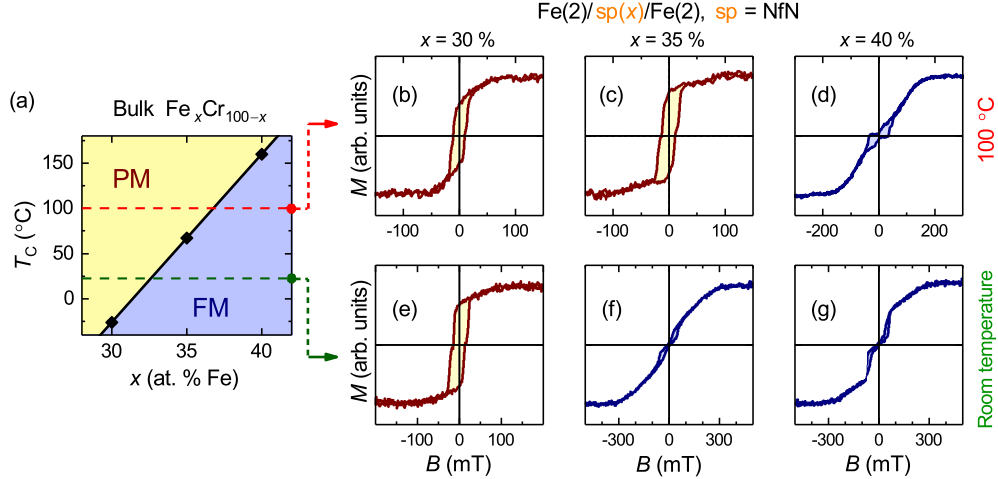}
\caption{(a) Linear interpolation of bulk Curie temperature vs. concentration $x$ for Fe$_x$Cr$_{100–x}$ binary alloy for $x$ = 30--40 at.~\% Fe  (data from ref.~\cite{r28}). M-B loops for NfN-samples with $x$ = 30, 35, and 40~\%, measured at 100~$^{\circ}$C (b,c,d) and room temperature (e,f,g).}
\label{fig_2}
\end{figure*}

The Fe/Cr/Fe system is known for its strong IEC when the Cr thickness is below approximately 3~nm, as illustrated in fig.~\ref{fig_1}(a), and vanishing IEC for spacers thicker than 3~nm. This well-established property is at the core of our design, which, in essence, achieves a rather abrupt change of the effective thickness of a specially designed composite spacer layer.

Classical RKKY-type Fe(2~nm)/Cr($d$)/Fe(2~nm) tri-layers with single-layer nonmagnetic (N) Cr spacers of thickness $d =$ 1--10~nm were fabricated as reference samples. These conventional N-type samples for $d < 3$~nm show the expected AFM interlayer coupling. The respective M-B loops have zero remnant magnetic moment $M_\text{r}$ and high saturation field $B_\text{s}$, illustrated in fig.~\ref{fig_1}(c) for $d = 1.5$~nm. The zero remnant magnetic moment corresponds to AFM ordering of the magnetic moments of the two Fe layers, induced by a rather strong IEC –-- up to 0.5~T is needed to align the magnetic moments of the Fe layers in parallel. The step-like shape of the M-B loop is due to an interplay of the RKKY interlayer exchange and the intrinsic anisotropy of the Fe layers (as detailed below and in Supplementary material A and C). Increasing $d$ reduces the saturation field $B_\text{s}$ and, at $d \geq 3$~nm, a single rectangular M-B loop is observed indicating vanishing IEC, with the outer Fe layers exchange-decoupled and switching independently. The shape of the M-B loop for $d = 10$ nm is taken as a reference for the structure ‘without IEC’. 
 
To demonstrate efficient thermal switching of IEC, the Fe/Cr/Fe tri-layer structure was modified to include a composite Cr/Fe$_x$Cr$_{100-x}$/Cr spacer, instead of the single-layer Cr spacer. Here, the layer of dilute ferromagnetic alloy Fe$_x$Cr$_{100-x}$, with $x =$ 30--40~at.~\%, is weakly ferromagnetic with a relatively low bulk Curie temperature, $T_\text{C} =$ 250--450~K \cite{r28}. In order to achieve an easy to detect AFM coupling between the outer Fe layers and, thereby, a clear illustration of the sought effect, a double composite spacer was used, with the specific structure of  Fe(2)/Cr(1.5)/Fe$_x$Cr$_{100-x}$(3)/Cr(1.5)/Fe$_x$Cr$_{100-x}$(3)/ Cr(1.5)/Fe(2), illustrated in fig.~\ref{fig_1}(b). The initial M-B data revealed no AFM alignment in any of the samples: rectangular M-B loops characteristic of structures without IEC were observed, such as that for the structure with $x = 35~\%$ in fig.~\ref{fig_1}(c). Clearly, the RKKY in the structure was too weak to transmit through the composite spacer and needed to be enhanced.
 
Additional experiments were performed at room temperature and confirmed a ferromagnetic state in 20-nm thick single-layer films of the Fe$_x$Cr$_{100-x}$ alloy with $x = 35$ and 40~at.~\%, but a paramagnetic state for $x = 30$~at.~\% (see also Suppl.~B for more details). This indicated that the Fe$_x$Cr$_{100-x}$ layers and the respective Fe$_x$Cr$_{100-x}$/Cr interfaces for the selected alloy compositions are at the cusp of ferromagnetic ordering and, hence, have near vanishing interface spin polarization. In order to enhance the spin-polarization at the Cr/Fe$_x$Cr$_{100-x}$ interfaces \cite{r29}, which is the key for RKKY, while maintaining the chosen spacer’s $T_\text{C}$, the ferromagnetic alloy layer was clad with ultrathin (0.25~nm) pure Fe layers, essentially making all six RKKY-active interfaces to be of type Fe/Cr. The resulting multilayer structures, with modified spacers, Fe(2)/Cr(1.5)/Fe(0.25)/Fe$_x$Cr$_{100-x}$(3)/Fe(0.25)/Cr(1.5)/ Fe(0.25)/Fe$_x$Cr$_{100-x}$(3)/Fe(0.25)/Cr(1.5)/Fe(2) (hereafter NfN refers to spacers with interface-enhanced f-layers), were found to be highly effective in establishing strong AFM IEC in the system, as shown by the magnetization data for the NfN-sample with $x = 35~\%$ in fig.~\ref{fig_1}(d). The observed M-B loops have zero remnant magnetic moment, much like the classical RKKY for conventional F/N/F (Fe(2)/Cr(1.5)/Fe(2) in fig.~\ref{fig_1}(c)).
 
Fe$_x$Cr$_{100-x}$ alloys of the concentration interval $x =$ 30-40~at.~\% Fe are known to have good interatomic solubility and the ferromagnetic-to-paramagnetic transition just above room temperature \cite{r28} (fig.~\ref{fig_2}(a); $T_\text{C}$ = 20--100~$^{\circ}$C) --- the range of interest for applications. The M-B loops of the NfN-samples with $x = 30$ and 35~\% measured at 100 $^{\circ}$C are of rectangular shape, with high remanence (fig.~\ref{fig_2}(b)--(d)). In contrast, the M-B loop for $x = 40~\%$ has zero remnant magnetic moment ($M_\text{r}$) and high saturation field ($B_\text{s}$). This indicates strong AFM coupling in the structure with $x = 40~\%$, whereas for $x = 30$ and 35~\% the outer Fe layers are essentially decoupled. At room temperature, the IEC-character of the M-B loops for the NfN-samples with $x = 30$ and 40~\% are the same as those at 100~$^{\circ}$C (fig.~\ref{fig_2}(e)--(g)), decoupled and AFM-coupled, respectively. In contrast, the M-B loop for the structure with $x = 35~\%$ completely changes its character, to AFM-type at room temperature with zero moment at zero field. This clearly shows that in the temperature interval of 20--100~$^{\circ}$C the spacer containing the spin-enhanced Fe$_{35}$Cr$_{65}$ layer undergoes a para-to-ferromagnetic transition (effective Curie temperature, $T_\text{C}^\text{eff}$, is within this interval). The measured behavior confirms the proposed switching mechanism illustrated in fig.~\ref{fig_1}(b).
 
\begin{figure}
\centering
\includegraphics{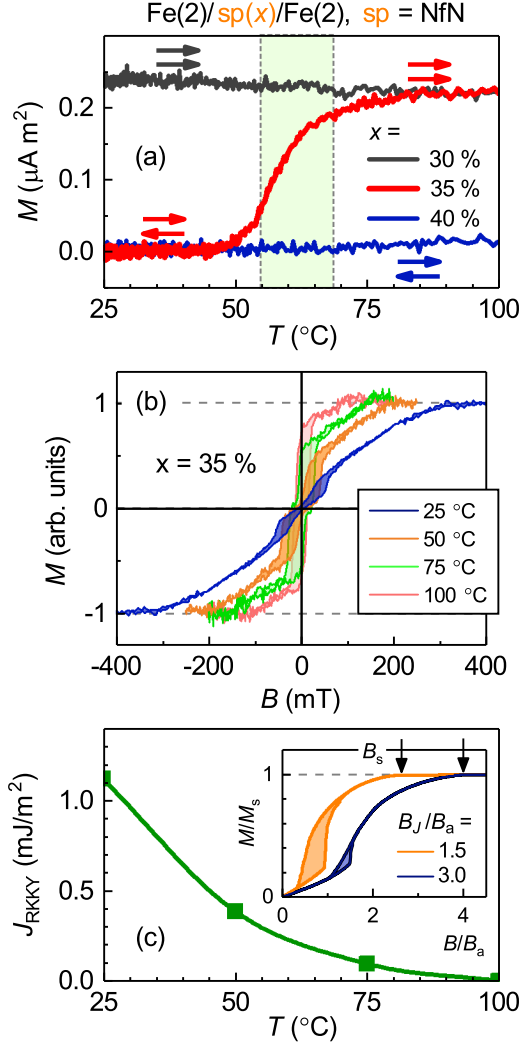}
\caption{(a) Magnetic moment vs. temperature for samples with $x = 30, 35\text{ and }40~\%$, measured on cool down with applied field of 1~mT. Arrows indicate the mutual orientation of the magnetic moments of the outer Fe layers. (b) Temperature evolution of the M-B loops for the structure with $x = 35~\%$. (c) Simulated temperature dependence of the RKKY exchange constant for NfN-sample with $x = 35~\%$; inset shows the calculated $M(B)$ for different representative strength of $J_\text{RKKY}$, with the vertical arrows indicating the corresponding saturation fields.}
\label{fig_3}
\end{figure}
 
The observed thermal switching between the parallel and antiparallel states of the above spin-valve-type structure should be attractive for device applications, provided a good switching performance can be achieved. In this regard, the temperature interval, within which the switching occurs, is the key characteristic.  $M$ vs. $T$ was measured in a weak applied field of 1~mT after saturating the sample into its parallel magnetic state at 110~$^{\circ}$C using the field of 100~mT (fig.~\ref{fig_3}(a)). Switching between the parallel (P) and antiparallel (AP) states of the NfN-structure with $x = 35~\%$, driven by the phase transition in the RKKY-IEC at the Curie point of the spacer, occurs in the interval of $\sim15~^{\circ}$C, which is quite narrow for a thermo-magnetic transition in a multilayer system. $M(T)$ for the same structure with $x = 30\text{ and }40~\%$ reveals no such transition, which is consistent with the expected behavior for these compositions, with either only AFM or no interlayer coupling present in the operating temperature range (20--100~$^{\circ}$C). These results show that the effective Curie point ($T_\text{C}^\text{eff}$) is easily tunable by choosing the appropriate Fe-concentration in the Fe-Cr layer of the spacer (f). The M-B loops for the NfN sample with $x = 35~\%$, recorded at different temperatures and shown in fig.~\ref{fig_3}(b), illustrate the thermo-magnetic transition in detail. As the temperature increases, the saturation field $B_\text{s}$ decreases, which is a reflection of a gradual suppression of the IEC. The remnant magnetic moment, on the other hand, has a much steeper temperature profile (red line in fig.~\ref{fig_3}(a)), indicating that the P to AP transition at $T_\text{C}^\text{eff}$ is more of a threshold type. 
 
Using the above data of fig.~\ref{fig_3}(b) for the temperature dependence of the hysteresis loop, one can obtain the temperature dependence of the effective RKKY-exchange constant (see Suppl.~C for details). Here we use a simple phenomenological model, with the RKKY coupling energy per unit area of bi-linear form, $W = J_\text{RKKY}(\mathbf{m}_1\cdot\mathbf{m}_2) = J_\text{RKKY}\cos\Delta\phi$,  where $\Delta\phi$ is the angle between the two outer Fe magnetic moments $\mathbf{m}_1$ and $\mathbf{m}_2$ \cite{r26,Kobler1992}, which is appropriate in our case of zero remanence magnetization at zero field. Including the suitable Zeeman and Fe-layer anisotropy terms and minimizing the full magnetic energy yields $M(B)$, shown in the inset to fig.~\ref{fig_3}(c). The simulated $M(B)$ closely resemble our experimental data shown in fig.~\ref{fig_3}(b) with all the main features explained by the interplay of RKKY, anisotropy, and Zeeman contributions: the saturation field and the coercive step, for example, shift to lower fields as the RKKY strength is reduced (at higher $T$), while the coercivity of the minor loop increases. The model shows that the saturation field is a sum of the effective RKKY and anisotropy fields, $B_\text{s}=B_j+B_\text{a}$, with the latter readily obtained from calibration measurements on single Fe-films of the same thickness and morphology (deposited under same conditions) as those used in the RKKY multilayers --- in our case $B_\text{a}=80$~mT. Subtracting $B_\text{a}$ and converting according to $J_\text{RKKY} = 2MB_jd$, yields the temperature dependence of the effective RKKY exchange coupling strength for the structure, shown in fig.~\ref{fig_3}(c). The saturation magnetization $\mu_0 M$ and thickness $d$ of the Fe layers were taken to be 2~T and 2~nm, respectively. The obtained $J_\text{RKKY}(T)$ shows that the RKKY coupling changes by an order of magnitude in the vicinity of the Curie transition in the spacer, thus demonstrating the great efficiency of thermal control of RKKY in our heterostructure. 

The following mechanism explains the observed thermal switching of the IEC in the multilayer. The indirect exchange interaction (RKKY) between the Fe and Fe-Cr layers is mediated by the conductance electrons in the Cr layers of the composite spacer. In this, the ultrathin Fe layers at the Cr/Fe-Cr interfaces provide a sufficient degree of spin polarization of scattered electrons for creating a coherent spin-density-wave state in the Cr layers. The exchange interaction within the weakly ferromagnetic Fe-Cr layers couples its two interfaces by direct exchange within the layer. Thus, the indirect exchange through the Cr layers and the direct exchange through the Fe-Cr layers form a serial sequence of interactions, providing the effective RKKY/direct-exchange interlayer coupling between the outer Fe layers (fig.~\ref{fig_1}(b)). With increasing temperature, above $T_\text{C}^\text{eff}$, the weakly ferromagnetic Fe-Cr layers undergo a FM-PM phase transition, which is rather sharp. As a result, the the direct exchange links within the Fe-Cr layers are suppressed and the effective outer Fe-Fe exchange is switched off. Only a small fraction of the available RKKY coupling ($<1~\%$ out of ~0.5~T in our experiment) is needed to rotate a relatively soft outer Fe layer (coercivity of about 0.01~T), so the resulting thermal transition is rather narrow.

Multiple optimization paths are straight forward, such as optimizing the choice of materials, compositional profiles, interfacial spin-enhancement, morphology (roughness), etc., should result in still better performance, likely with sub-10-K RKKY-Curie-transition widths. This, however, goes beyond the scope of this letter, focused on demonstrating the effect of thermal RKKY-switching in magnetic multilayer systems. 

Temperature dependent IEC was reported for other material systems, based on \emph{direct exchange} via weakly ferromagnetic spacers between two strongly ferromagnetic outer layers \cite{Demirtas_PRB,r30,r31}. However, the direct-exchange designs exhibit strong proximity effects and associated limitations on the multilayer design. In contrast, efficient thermal control of RKKY makes it possible to have either AFM of FM ground state in the structure, which is not possible using only direct exchange.

\section{Conclusion}

In summary, we report on a new concept for thermo-magnetic switching in multilayers, exploiting a combination of indirect and direct exchange. We demonstrate such on/off IEC switching in a Fe/Cr/FeCr-based system and obtain RKKY switching as sharp as 10--20~K, essentially in any desired temperature range. High design tuneability in the physical parameter space of field-temperature-magnetization, along with availability of a  wide choice of materials, make AFM-RKKY or no-RKKY ground states easily obtainable, unlike that in synthetic antiferro- or ferri-magnets. We believe that the demonstrated effect of thermally controlled indirect exchange coupling adds a new degree of freedom to designing future spin-electronic devices, such as memory\cite{Prejbeanu_2013} and oscillators \cite{Kadigrobov_2010}.




\acknowledgments{
Support from the Swedish Research Council (VR Grant No. 2014-4548) and the Swedish Stiftelse Olle Engkvist Byggm\"astare is gratefully acknowledged.}

\bibliography{References}

\end{document}